\documentclass[a4paper,fleqn]{cas-sc}

\usepackage[utf8]{inputenc}
\usepackage[T1]{fontenc}
\usepackage{graphicx}
\usepackage[square,numbers,sort&compress]{natbib}

\usepackage{dcolumn}

\usepackage{xcolor}

\def\tsc#1{\csdef{#1}{\textsc{\lowercase{#1}}\xspace}}
\tsc{WGM}
\tsc{QE}
\tsc{EP}
\tsc{PMS}
\tsc{BEC}
\tsc{DE}

\ExplSyntaxOn
\keys_set:nn { stm / mktitle } { nologo }
\ExplSyntaxOff

\usepackage{amsthm}
\theoremstyle{plain}
\newtheorem{remark}{Remark}

\begin{document}


\shorttitle{Power Laws, the Price Model, and the Pareto type-2 Distribution}
\title[mode = title]{Power Laws, the Price Model, and the Pareto type-2 Distribution}

\shortauthors{Siudem, Nowak, Gagolewski}

\author[1]{Grzegorz Siudem}[orcid=0000-0002-9391-6477]
\ead{grzegorz.siudem@pw.edu.pl}
\ead[URL]{http://if.pw.edu.pl/~siudem}
\cormark[1]
\cortext[cor1]{Corresponding author}
\credit{%
Conceptualisation of this study,
Formal analysis,
Methodology,
Writing -- Original draft preparation%
}

\author[1]{Przemysław Nowak}
\credit{%
Investigation,
Formal analysis,
Validation%
}

\author[2,3]{Marek Gagolewski}[orcid=0000-0003-0637-6028] 
\ead{m.gagolewski@deakin.edu.au}
\ead[url]{https://www.gagolewski.com}
\credit{%
Conceptualisation of this study,
Formal analysis,
Methodology,
Software,
Visualisation,
Writing -- Original draft preparation%
}

\address[1]{Warsaw University of Technology, Faculty of Physics,
ul.~Koszykowa 75, 00-662 Warsaw, Poland}

\address[2]{Deakin University, School of IT, Geelong, VIC 3220, Australia}

\address[3]{Systems Research Institute, Polish Academy of Sciences,
ul.~Newelska 6, 01-447 Warsaw, Poland}
%

\date{\today}

\begin{abstract}
We consider a version of D.~Price's model for the growth of a bibliographic
network, where in each iteration, a constant number of citations is randomly
allocated according to a weighted combination of the accidental (uniformly
distributed) and the preferential (rich-get-richer) rule.
Instead of relying on the typical master equation approach,
we formulate and solve this problem in terms of the rank-size distribution.
We show that, asymptotically, such a process leads to a Pareto-type 2
distribution with a new, appealingly interpretable parametrisation. We prove
that the solution to the Price model expressed in terms of the rank-size
distribution coincides with the expected values of order statistics in an
independent Paretian sample.
An empirical analysis of a large repository of academic papers
yields a good fit not only in the tail of the distribution (as it is usually
the case in the power law-like framework), but also across a significantly
larger fraction of the data domain.
\end{abstract}


\maketitle


\section{Introduction}

Citing papers is a common way to express appreciation of someone else's research
or to acknowledge the relevance thereof with regard to our own results
\cite{thelwall2021new}. Regardless of the true  motivation
\cite{lyu2021classification}, the what-to-cite decisions of individual
authors can be averaged over the whole bibliographic networks.
This way, we may study the underlying mechanisms governing the
emergence of particular citation distributions.
Understanding these is of interest not only to network science,
but also to the entire research community, especially bearing in mind that
citation counts are widely used as a (controversial \cite{Gagolewski2013:fair,dorogovtsev2015ranking}) proxy for articles',
authors', and journals' impact.

One such mechanism is known as rich-get-richer,
success-breads-success \cite{Price1963-book},
the Matthew effect \cite{Merton1968},
or the preferential attachment rule \cite{perc2014matthew}.
It assumes that highly cited papers are most likely to receive even more
citations in the future (for the possible bibliometric applications of this rule
see \cite{IonescuChopard,Zogala-Siudem2016}).
Yet, recent research \cite{sinatra2016quantifying,janosov2020success}
into the origins of success in science and beyond highlights the role of other
factors -- such as chance. Some  \cite{pluchino2018talent,pluchino2019exploring}
claim that luck contributes more than the beneficiaries are eager to admit.

This  leads to the question: how to measure the level of randomness in a
citation network? It is noted in \cite{heesen2017academic} that when
citations follow the rich-get-richer rule, it is virtually impossible to
distinguish between the merit- or non-merit-driven motivations.
The literature knows
several other approaches to modelling the structure of the networks in general
\cite{kharel2021degree}, and citation networks in particular, including those
based on recursive searching, node duplicating, and fitness factors
\cite{Golosovsky2017,Steinbock2017,Golosovsky2018,%
steinbock2019analytical,steinbock2019analytical2}.
They usually employ a significant number of parameters which reduces
their interpretability. What is more, they do not allow for a direct measurement
of the level of accidentality, which is our intention herein.

For these reasons, in this paper, we shall consider a classical D.~Price's
model \cite{price1976general} which explicitly combines the two said mechanisms.
We shall discuss a few methods for identifying the preferential-to-accidental
ratio from real-world data.
The Price model was studied in, amongst others,
\cite[Chap.~14]{newman2018networks}
and frequently appears in the literature under different names and modifications
\cite{liu2002connectivity,bedogne2006complex,shao2006growing,peterson2010nonuniversal,Neda2017,PNAS2020}
and in different contexts: e.g., analysis of resistance to random failures and
intentional attacks in complex networks \cite{liu2002connectivity}
or computation of longest paths in random graphs \cite{evans2020longest}.

The most typical approach (e.g., \cite{peterson2010nonuniversal})
to deriving the citation distribution and thus the preferential-to-accidental
ratio in the Price model is via master equations (see \cite{Dorogovtsev2000}).
In this work, however, we shall apply the rank-size (order statistics)
approach which was inspired by our earlier work \cite{PNAS2020}, where
we studied citation vectors of individual scientists, i.e., in small scale, but
with similar accidental and preferential contributors.
Here we shall modify the model's boundary conditions so that we can focus
on papers which have obtained a sufficiently large number of citations
(e.g., 1, to avoid problems with computing and drawing on the log scale).
Moreover, we shall consider citation networks in their entirety, i.e.,
study the model's asymptotic behaviour.
We shall show that this leads to a well-known generalisation of the
Pareto distribution (\cite{arnold2015:pareto,bourguignon2016new,figueira2011gompertz}) albeit
with a new, very appealing parametrisation. This way we make an interesting
addition to the list \cite{arnold2015:pareto} of processes from which
such a distribution emerges. Revealing the connection between the Price model
and the Pareto-type 2 distribution will allow us to estimate the
preferential-to-accidental ratio using some more statistically reliable
methods than those previously mentioned in the literature.

\medskip
This work is set out as follows.
In Section~\ref{sec:rank} we introduce the rank distribution approach to the
preferential-accidental attachment process, which allows us to establish the
relation between the Price model and the Pareto-type 2 distribution.
In Section~\ref{sec:estimators} we discuss three methods for estimating
the model parameters from data and quantify their usability.
Further, in Section~\ref{sec:data} we apply them on the DBLP
\cite{ArnetMiner} repository of computer science papers.
Lastly, in Section~\ref{sec:final} we discuss the implications of our
findings and propose some directions for future research.


\section{Rank distribution approach to Price's model}\label{sec:rank}

Let us consider a process where in every time step, a citation network grows
by one new paper with $\delta\ge 0$ initial\footnote{%
Note that in \cite{PNAS2020}, where we considered a single author's record
and not the whole citation network, we assumed $\delta=0$. Our more general
setting might now mean, for example, that we become interested in a paper's
existence not immediately after it has been published, but only when it has
reached some level of maturity/recognisability/attention-worthiness.
Furthermore, we abstract from the many possible origins of the citations that
have emerged before the paper has appeared in the simulation.}
citations.
This new paper features $m-\delta$ references to the articles that already exist
in the system:

\begin{enumerate}
\item
$a=(1-\rho)(m-\delta)$ citations are allocated completely at random,
\item
$p=\rho(m-\delta)$ citations follow the preferential attachment
(i.e., rich-get-richer rule; compare \cite{perc2014matthew}) rule,
\end{enumerate}

\noindent
with $a+p=m-\delta$ and $\rho\in(0,1)$ representing
the extent to which the rich-get-richer rule dominates over pure luck.
Note that $m> \delta$ gives the total number of citations
added into the system in every time step.

\subsection{Exact solution}

Let $X_k(t)$ denote the number of citations to the $k$-th most cited paper
at a time step $t$.
We assume that $X_k(k-1)=\delta$ for every $k$, i.e.,
the $k$-th publication enters the system with $\delta\in[0,\, m)$ citations.

According to the above description, the number of citations
grows between the iterations as follows:

\begin{equation}\label{eq:3dsi}
X_k(t) = \underbrace{X_k(t-1)}_{\mathrm{previous\;value}} + \underbrace{\frac{a}{t}}_{\mathrm{accidental\;income}} + \underbrace{p \,\frac{X_k(t-1) + \frac{a}{t}}{\delta+(t-1)m + a}}_{\mathrm{preferential\;income}}=X_k(t-1) \frac{t}{t+\phi}+\frac{a}{t+\phi},
\end{equation}

\noindent
where $a$ and $p$ were defined above and for brevity we denote:
\begin{equation}\label{eq:phi}
    \phi=-\frac{p}{m}
=\rho\left(\frac{\delta}{m}-1\right).
\end{equation}

Let us iterate Eq.~(\ref{eq:3dsi}) in a similar manner as we did in
\cite{PNAS2020}:
\begin{align*}
X_k(t)=&X_k(t-2) \frac{t-1}{t-1+\phi}\frac{t}{t+\phi}+\frac{a}{t-1+\phi} \frac{t}{t+\phi}+\frac{a}{t+\phi} =\underbrace{X_k(k-1)}_{=\delta} \frac{k\cdot (k+1) \cdot \dots \cdot t}{(k+\phi)\cdot (k+1+\phi) \cdot \dots \cdot (t+\phi)}+\\
&+\frac{a\cdot(k+1) \cdot \dots \cdot t}{(k+\phi)\cdot (k+1+\phi) \cdot \dots \cdot (t+\phi)}+\frac{a\cdot(k+2) \cdot \dots \cdot t}{ (k+1+\phi) \cdot \dots \cdot (t+\phi)}+\dots +\frac{at}{(t-1+\phi)(t+\phi)}+\frac{a}{t+\phi}.
\end{align*}
This can be simplified further thanks to the notion of the Euler gamma
function:
\begin{align*}
    X_k(t)=\delta \frac{\Gamma(t+1)\Gamma(k+\phi)}{\Gamma(k)\Gamma(t+1+\phi)}+a\sum_{\ell=k}^t \frac{\Gamma(t+1)\Gamma(\ell+\phi)}{\Gamma(\ell+1)\Gamma(t+1+\phi)}=
    \delta \frac{\Gamma(t+1)\Gamma(k+\phi)}{\Gamma(k)\Gamma(t+1+\phi)}-\frac{a}{\phi}\left( \frac{\Gamma(t+1)\Gamma(k+\phi)}{\Gamma(k+1)\Gamma(t+1+\phi)}-1\right),
\end{align*}
which leads to:
\begin{equation}\label{eq:ikska}
X_k(t)
=\left(\delta+m\frac{1-\rho}{\rho}\right)\frac{\Gamma(t+1)}{\Gamma(k)}
\frac{\Gamma(k+\phi)}{\Gamma(t+1+\phi)} - m\frac{1-\rho}{\rho}.
\end{equation}

\medskip
Let us stress that, in each iteration $t$, the above formula preserves
the average number of citations, i.e., it holds:

\begin{equation}\label{eq:sum}
\frac{1}{t} \sum_{k=1}^t X_k(t) = m.
\end{equation}

\subsection{Limiting case}

Let us recall the Gautschi inequality (see Eq.~(7) in \cite{Gautschi})
which states that for any $k\geq 0$ and $\psi\in [-1,\,0]$, it holds:

\begin{equation}
k^{\psi}\leq \frac{\Gamma(k+\psi)}{\Gamma(k)} \leq (k-1)^{\psi}.
\end{equation}

Eq.~\eqref{eq:ikska} can be rewritten as a function of $y=k/t$.
Applying the above inequality yields in the limit as $t\rightarrow\infty$:

\begin{eqnarray}\label{eq:cquant}
X(y)&:=&\lim_{t\rightarrow\infty}X_{yt}(t)=
m\frac{\rho-1}{\rho}+\frac{m+\delta\rho-\rho m}{\rho}y^{\phi}=
\left(\delta + m\frac{1-\rho}{\rho}\right) y^{-\rho(1-\delta/m)}
- m\frac{1-\rho}{\rho}.
\end{eqnarray}

\noindent
The inverse of $X$, denoted $S=X^{-1}$, is given for $x\ge \delta$ by:

\begin{eqnarray}
S(x)=\left(\frac{x +m/\rho-m}{\delta +m/\rho- m}\right)^{1/\phi}
 =\left(1+\frac{x-\delta}{m/\rho-(m-\delta)}\right)^{1/\phi}
 =
\left(\frac{m-\rho(m-\delta)}{\rho x + (1-\rho)m}\right)^{\frac{m}{\rho(m-\delta)}}.\label{eq:ccdf}
\end{eqnarray}

\noindent
and of course $S(x)=1$ for $x < \delta$.

It is a strictly decreasing continuous function onto $(0, 1]$.
Hence, we can treat it as some complementary cumulative distribution
function (CCDF; also known as the tail distribution or the survival function).
Generally, if a random variable $Z$ has a CCDF $S$, then $S(x)=\Pr(Z>x)$.
Its inverse $X=S^{-1}$ is referred to as the complementary quantile function.

From now on shall refer to $S$ given by Eq.~\eqref{eq:ccdf}
as the CCDF of the \textit{$\delta$-truncated Price distribution}
with parameters $m>\delta\ge 0$ and $\rho$.
This way we honour the contributions of D.~Price who, as we mentioned
in the introduction, studied a similar model in \cite{price1976general}.

\subsection{Price meets Pareto}

Let us re-express the CCDF given by Eq.~\eqref{eq:ccdf}:

\begin{equation}\label{eq:pareto2}
S(x) = \left(
1+\frac{x-\delta}{\lambda}
\right)^{-\alpha}
=
\left(
\frac{\lambda}{x-\delta+\lambda}
\right)^{\alpha}
\qquad(x\ge \delta),
\end{equation}

\noindent
where:

\begin{equation}
\alpha=\frac{m}{\rho (m-\delta)}   \;\;(\alpha > 1)\;\;\;\mathrm{and}\;\;\;\lambda=(m-\delta) (\alpha-1)   \;\; (\lambda>0).
\end{equation}

\noindent
This is nothing else than the standard parametrisation of the
Pareto-type 2 distribution (e.g., \cite{arnold2015:pareto}).
In some sources, it is also called the Lomax distribution shifted by $\delta$.

Hence, the above derivations form an interesting addition
to the catalogue of processes from which the Pareto-type~2 distribution
emerges; see \cite{arnold2015:pareto} for an overview of other ones.

\bigskip
Furthermore, let us emphasise that in the special case of
$\rho=1$ (purely preferential attachment),
Eq.~\eqref{eq:ccdf} reduces to:

\begin{equation}\label{eq:pareto21}
S_1(x) = \left(
\frac{x}{\delta}
\right)^{-\alpha}
\qquad(x\ge \delta),
\end{equation}

\noindent
with: $\alpha=m/(m-\delta)$ and $\alpha>1$, i.e., Eq.~\eqref{eq:pareto2} with $\lambda=\delta$.
This is known in the literature as the (type-1) Pareto distribution;
see \cite{arnold2015:pareto}.

\subsection{Back to the ranks}

Despite our model's stemming from a deterministic setting,
its asymptotic expansion gives a description of a whole population
of papers, from which we can then pick items at random.
It might be interesting to see how the distribution we have just derived
relates to Eq.~\eqref{eq:ikska}.

First, however, let us note (e.g., \cite[Eq.~21.9]{Billingsley}) that
the expected value of a random variable $X\ge \delta\ge 0$
with a continuous CCDF $S$ is given by the tail-sum formula:
\begin{equation}
\mathbb{E}\,[X] = \int_0^\infty S(x)\,dx = \delta+\int_\delta^\infty S(x)\,dx.
\end{equation}

This is equivalent to:
\begin{equation}
\mathbb{E}\,[X] = \int_0^1 S^{-1}(y)\,dy.
\end{equation}

In our case, where $S^{-1}$ is given by Eq.~\eqref{eq:cquant},
it holds $\mathbb{E}\,[X] = m$, which is consistent with our model's assumptions:
a sample of randomly selected papers will have $m$
citations on average (compare Eq. (\ref{eq:sum})).

\medskip
Further, given an independent, identically distributed sample of random variables
$X_1,\dots,X_n$ following a CCDF $S$,
let $X_{r:n}$ denote the $r$-th order statistic, i.e., the $r$-th
smallest value therein. Hence, $X_{1:n}$ is the sample minimum
and $X_{n:n}$ is the maximum.

Let us derive the formula for $\mathbb{E}\,[X_{r:n}]$.
Of course, from Eq.~(4.5.1) in \cite{DavidNagaraja2003:orderstatistics}
we get that for large $n$ it approximately holds:
\begin{equation}
\mathbb{E}\,[X_{r:n}] \simeq 1-S^{-1}\left(\frac{r}{n+1}\right).
\end{equation}
But we can do better than this: let us provide an exact formula.
Namely, knowing that the CCDF of the $r$-th order statistic is (e.g.,
\cite{DavidNagaraja2003:orderstatistics}):

\begin{equation}
S_{r:n}(x) = 1-\sum_{j=r}^{n}{\binom{n}{j}}(1-S(x))^{j}(S(x))^{n-j},
\end{equation}

\noindent
we obtain:

\begin{eqnarray}
\mathbb{E}\,[X_{r:n}]
&=& \int_0^\infty S_{r:n}(x)\,dx
=\delta+\int_{\delta}^\infty\left[\sum_{j=0}^n\binom{n}{j}\left(1-S(x)\right)^jS^{n-j}(x)-\sum_{j=r}^n\binom{n}{j}\left(1-S(x)\right)^jS^{n-j}(x)\right]dx
\nonumber
\\
&=&\delta+\sum_{j=0}^{r-1}\binom{n}{j}\int_{\delta}^\infty\left(1-S(x)\right)^jS^{n-j}(x)dx.
\end{eqnarray}

\noindent
Applying the binomial theorem
to $\left(1-S(x)\right)^j$ and performing some elementary integration,
we get:

\begin{eqnarray}
\mathbb{E}\left[X_{r:n} \right]
&=&\delta+(m-\delta)\sum_{j=0}^{r-1}\Bigg[\binom{n}{j}\sum_{\ell=0}^j\binom{j}{\ell}(-1)^\ell\tfrac{\delta-m+m/\rho}{\delta-m+(n-j+\ell)m/\rho}\Bigg]
\nonumber
\\
&=& m\frac{\rho-1}{\rho}+(-1)^r\frac{m+\rho\delta-\rho m}{\rho}
\frac{\Gamma(n+1)}{\Gamma(n-r+1)}
\frac{\Gamma\left(-(n+\phi)\right)}{\Gamma\left(-(n-r+\phi)\right)}.
\end{eqnarray}

\noindent
Let us also note that:

\begin{equation}
\frac{\Gamma\left(-(n+\phi)\right)}{\Gamma\left(-(n-r+\phi)\right)}=(-1)^r\frac{\Gamma(n-r+1+\phi)}{\Gamma(n+1+\phi)}.
\end{equation}

\noindent
Combining the above yields:

\begin{equation}\label{eq:orderfinal}
\mathbb{E}\left[X_{r:n} \right]=m\frac{\rho-1}{\rho}+\frac{m+\rho\delta-\rho m}{\rho}\frac{\Gamma(n+1)}{\Gamma(n-r+1)} \frac{\Gamma(n-r+1+\phi)}{\Gamma(n+1+\phi)}.
\end{equation}

Equation~\eqref{eq:orderfinal} is identical to Eq.~\eqref{eq:ikska}
with $k=n-r+1$ and $t=n$.
This observation strengthens the rationale behind
our asymptotic expansion even further: the rank-size approach
gives the expected values of the order statistics
from any finite sample therefrom, including a very small one.

\subsection{Price meets power laws}\label{sec:powerlaw}

Newman in \cite{Newman2005} refers to the CCDF of the Pareto-type 1 distribution
given by Eq.~\eqref{eq:pareto21}
as a power law distribution (compare Eq.~(4) therein). This is due to the fact
that, when plotted on a double log-scale,  $S_1(x)$ is a straight line with
slope $-\alpha$.

More generally, i.e., for any $\rho$, the Pareto-type 2 distribution
follows a power law at least in its tail.
Namely, the plot of $x\mapsto S(x)$ given by Eq.~\eqref{eq:pareto2}
on a double logarithmic scale looks like the plot of:

\begin{equation}
\ell \mapsto (-\alpha \log(\exp(\ell)-\delta+\lambda)+\alpha \log(\lambda))
\stackrel{\ell \text{ large}}{\simeq} -\alpha \ell + \alpha\log\lambda.
\end{equation}

\begin{figure}[b!]
\centering
\includegraphics[width=0.5\textwidth]{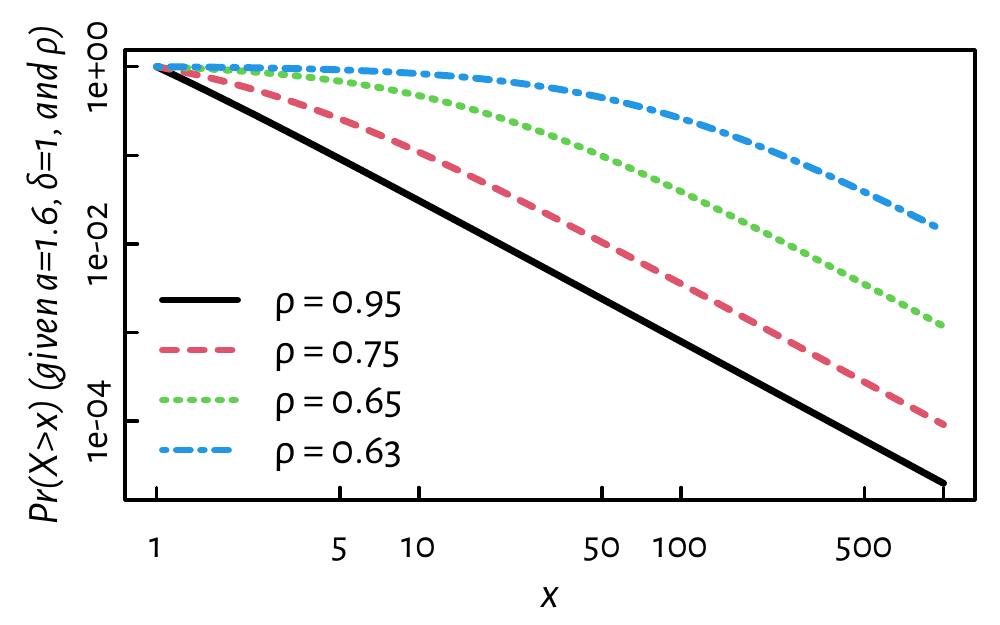}
\caption{\label{Fig:fixed_alpha}
CCDF of the Pareto-type 2 distribution
for fixed $\alpha=1.6$, $\delta=1$,
and different
$\lambda=\frac{\delta(\alpha-1)}{\rho\alpha-1}$
for given $\rho$s; note the double log-scale.
The tail of each distribution follows the power law with the same
exponent.
}
\end{figure}

\noindent
See Figure~\ref{Fig:fixed_alpha} for an illustration.
On a side note, if the slope of the CCDF's tail is $-\alpha$,
then the slope the probability density function's tail is $-\alpha+1$.

\bigskip
It is not rare in the data analysis practice
to consider only a truncated version of a given dataset,
and focus solely on its right tail; compare \cite{evt,Newman2005}.
This is particularly the case when we are merely interested in describing
the behaviour of ``significant'' objects in our database (excess),
e.g., highly-cited papers.

%
%

For a fixed $\alpha$, $\lambda$, and $\delta$, let $S$
be the CCDF of the corresponding Pareto-type 2 distribution,
$\Pr(X>x)$.
Consider any threshold $t>\delta$.
Then, the CCDF of the truncated distribution,
$\Pr(X>x|X>t)$ is given by:

\begin{equation}
S|_{t}(x) = \frac{S(x)}{S(t)}
= \left(
\frac{t-\delta+\lambda}{x-\delta+\lambda}
\right)^{\alpha}
= \left(\frac{\lambda'}{x-\delta'+\lambda'}
\right)^{\alpha}
\qquad(x\ge t).
\end{equation}

\noindent
It is thus still a Pareto-type 2 distribution
with the same $\alpha$,
but with different $\lambda'=t-\delta+\lambda$ and $\delta'=t$.

The corresponding $m'$ and $\rho'$ can also be recreated:

\begin{eqnarray}\label{eq:recreatedthreshold}
m' & = & \frac{t\alpha-\delta+\lambda}{\alpha-1} = m+(t-\delta)\frac{\alpha}{\alpha-1},\\
\rho' & = & \frac{t\alpha-\delta+\lambda}{\alpha (t-\delta+\lambda)}.
\end{eqnarray}

\begin{figure}[h!]
\centering
\includegraphics[width=0.5\textwidth]{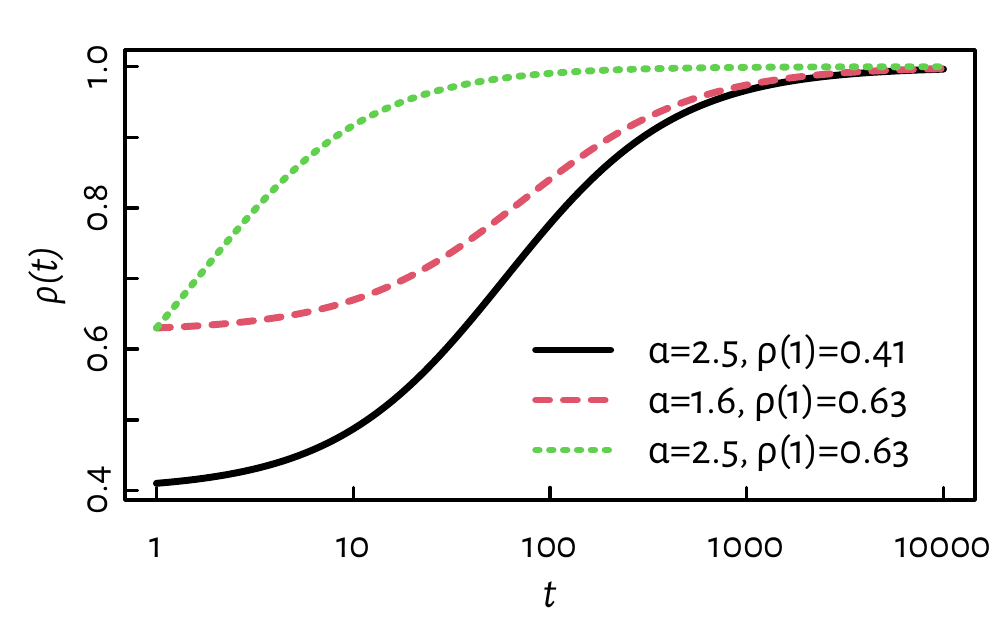}
\caption{\label{Fig:rho_threshold}
$\rho$ as a function of the truncation threshold $t$
for different
fixed $\alpha$s and initial $\rho$s at $\delta=1$.
It holds $\rho(t)\to 1$ as $t\to\infty$.
}
\end{figure}

\noindent
The linear increase of $m'$ (as a function of $t$) is unsurprising.
What \textit{is} particularly remarkable, though, is that the ratio of
preferentially to accidentally allocated citations increases
as the cutoff threshold increases;
see Figure~\ref{Fig:rho_threshold} for an illustration.

\normalcolor

\subsection{Closer to real-world data: Discretisation}

Typically, real-world empirical data are not continuous  
(for instance: raw citation counts). We should therefore study
a discretised version of our distribution as well.

Let us assume that data come 
from the original distribution, $X$, but what we observe is its
truncated version, $\lfloor X\rfloor$,
being the greatest integer less than or equal to $X$.
The CCDF of such a random variable is simply given by:

\begin{equation}
\Pr(\lfloor X\rfloor>x) = \Pr(X>\lfloor x\rfloor+1) = S(\lfloor x\rfloor+1).
\end{equation}

\begin{remark}
On a side note, in the limit as $N\to\infty$, for $\delta=0$,
the master equation approach (see \cite{Dorogovtsev2000})
applied to the same random process as above, yields the probability
mass function $\underline{p}(x)$ for any $x\in\mathbb{N}_0$ given by:

\begin{equation}\label{eq:Pk}
    \underline{p}(x)
     = \frac{1}{m + 1-\rho m} \, \frac{\Gamma(\frac{m}{\rho}- m + 1+\frac{1}{\rho})}{\Gamma(\frac{m}{\rho}- m)}
     \frac{\Gamma(x+\frac{m}{\rho}-m)}{\Gamma(x+\frac{m}{\rho}-m + 1 + \frac{1}{\rho})}= \frac{1}{m+1-\rho m} \frac{(m/\rho-m)_{1+1/\rho}}{(x+m/\rho-m)_{1+1/\rho}},
\end{equation}

\noindent
where $(a)_x=\Gamma(a+x)/\Gamma(a)$ denotes the Pochhammer symbol. Let us compare the distribution-domain results via the master equation approach, with the  CCDF-domain rank approach considered in this paper.
For this purpose, let us compute the CCDF from the above distribution.
And thus, it may be shown that for any integer $k$:

\begin{equation}
\sum_{\ell=0}^k \underline{p}(\ell)=1- 
 \frac{\Gamma \left(-m+\frac{m}{\rho}+\frac{1}{\rho }\right)}{\Gamma \left(-m+\frac{m}{\rho}\right)}
 \frac{ \Gamma \left(k-m+\frac{m}{\rho }+1\right)}{\Gamma \left(k-m+\frac{m}{\rho}+\frac{1}{\rho }+1\right)}.
\end{equation}

\noindent
Hence, the complementary cumulative distribution function
of this distribution is defined for any $x\ge 0$ as:

\begin{equation}\label{eq:CPk}
\underline{S}(x)=\sum_{\ell=\lfloor x\rfloor+1}^\infty p(\ell)=\frac{(m/\rho-m)_{1/\rho}}{(\lfloor x\rfloor+1+m/\rho-m)_{1/\rho}}.
\end{equation}

\noindent
By the aforementioned Gautschi inequality, our discretisation leads to:

\begin{equation}
S\left(\lfloor x\rfloor + 1 \Big| \delta=0\right)
=
\frac{(m/\rho-m)^{1/\rho}}{(\lfloor x\rfloor + 1+m/\rho-m)^{1/\rho}} \simeq \underline{S}(x),
\end{equation}

\noindent
which gives a nice correspondence between the master equation-
and the rank distribution-approach.
\end{remark}


\section{Estimating model parameters}\label{sec:estimators}

Fitting different heavy-tailed distributions
to real-world data is quite commonly exercised in the complex systems practice.
In particular, \cite{Newman2005,ClausetShaliziNewman2009} feature
a comprehensive discussion related to the fitting of the power
law distributions.

\subsection{All parameters are interpretable}

The model we have landed at is in fact a Pareto-type 2 distribution,
which by itself has of course been studied extensively;
see \cite{arnold2015:pareto}.
However, in this paper we have arrived at a non-classic parametrisation thereof:
all its parameters are easily interpretable. Namely,
$\delta$ gives the cut-off (distribution shift),
$m$ is the expected number of citations of a randomly selected paper
(and also the number of citations allocated per iteration),
whereas $\rho$ gives the ratio of preferentially to accidentally attached
citations.

The truncation threshold $\delta\ge 0$ is usually
known a priori; we therefore consider it fixed.
For instance, we might only be recording the incidence
of extreme values ($\Pr(X > x | X > \delta)$; compare \cite{evt}).
Moreover, if data modellers wish to investigate
only what is happening in the tail of the distribution,
they trim the observations themselves
(compare, e.g., \cite{Newman2005,ClausetShaliziNewman2009}
where only large-enough cities,
the richest Americans, the most popular surnames, etc., are
taken into account in the analysis).
Otherwise, just like in the case of the power law distribution in \cite{Newman2005}, we can estimate it by simply taking the smallest value
in the sample.

We should thus be interested in studying the properties of various basic
estimators of $m$ and $\rho$ so that we can make sure that we are able to
fit our model reliably to empirical data.

\subsection{Different estimators of $m$ and $\rho$}

From a continuous CCDF we can easily
obtain the underlying probability density function:

\begin{equation}
f(x|m,\rho)=\frac{d(1-S(x|m,\rho))}{dx}.
\end{equation}

\noindent
This enables us to compute the maximum likelihood estimator (MLE) of the
model parameters, being the solution to:

\begin{equation}
\max_{m,\rho} \sum_{i=1}^n \log f(X_i|m,\rho),
\end{equation}

\noindent
However, instead of solving directly with respect to $m$ and $\rho$,
we can consider the well-known, computationally less complicated maximum likelihood
estimator of $\lambda$, which is determined by numerically solving
the following nonlinear equation:

\begin{equation}
\frac{1}{n} \sum_{i=1}^n \log(x_i+\lambda-\delta)+ 1 - \log \lambda
- \frac{n}{\lambda \sum_{i=1}^n (x_i+\lambda-\delta)^{-1}}=0.
\end{equation}

\noindent
Then we get the MLE of $\alpha$:

\begin{equation}
\alpha=\left(\frac{1}{n} \sum_{i=1}^n \log(x_i+\lambda-\delta)-\log\lambda\right)^{-1}.
\end{equation}

\noindent
From this we obtain:

\begin{eqnarray}
m&=&\frac{\lambda}{\alpha-1}+\delta,\\
\rho&=&\frac{m}{\alpha(m-\delta)}.
\end{eqnarray}

\bigskip
For the sake of comparison,
we will consider two other estimators based on the empirical CCDF,

\begin{equation}
\hat{S}(x|X_1,\dots,X_n)=\frac{|\{i: X_i>x\}|}{n}.
\end{equation}

\noindent
The first estimator minimises the maximal difference between the empirical and
the theoretical CCDF (from now on called SUP):

\begin{equation}
\min_{m,\rho} \max_{x\in\{X_1,\dots,X_n\}}
\left|\hat{S}(x|X_1,\dots,X_n)-S(x|m,\rho)\right|,
\end{equation}

\noindent
and the second one minimises the sum of  squared errors  (SSE) at all
points of discontinuity of the empirical CCDF:

\begin{equation}
\min_{m,\rho} \sum_{x\in\{X_1,\dots,X_n\}}
\left(\hat{S}(x|X_1,\dots,X_n)-S(x|m,\rho)\right)^2.
\end{equation}

\noindent
The estimators will be computed using the Nelder--Mead method
with a logarithmic barrier
(via \texttt{constrOptim} in R) to enforce the constraints on  $m$ and $\rho$.
We observed that if the algorithm starts from $m$ equal to
the 90\% trimmed mean and $\rho=0.5$, then it always converges nicely.

\bigskip
Similarly, we will also consider the MLE, SSE, and SUP estimators
for the discretised version of the original random variable, $\lfloor X\rfloor$,
with the MLE being the (numerical) solution to:

\begin{equation}
\max_{m,\rho} \sum_{i=1}^n \log p(X_i|m,\rho),
\end{equation}

\noindent
where the probability mass function $p$ is given by
$p(x)=S(\lfloor x\rfloor)-S(\lfloor x\rfloor+1)$. Note that this time we are
optimising the likelihood directly with respect to $m$ and $\rho$.

\subsection{Assessing estimator quality}

Thanks to the principle of inverse transform sampling
and the exact formula for the inverse
of the CCDF given by Eq.~\eqref{eq:cquant}, i.e., $S^{-1}$, we can easily
generate realisations of independent samples like $X_1,\dots,X_n$
following our distribution for specific $\delta$, $m$, and $\rho$.
Namely, we set $X_i=S^{-1}(U_i)$, where $U_i$ is uniformly distributed
on the unit interval.

To study the quality of the estimators via the Monte Carlo approach,
we need to generate many samples, where each sample consists
of $n\in\{1{,}000, 100{,}000\}$ observations from the Pareto-type 2
distribution with $\delta=1$, $m=25$, and a range of $\rho$s from
the set $\{0.5, 0.75, 0.9\}$.

From the evaluation perspective and for the purpose of this study,
the choice of $\delta$ is not really
important, because we have already observed that a truncated Pareto-type 2
distribution is still a Pareto-type 2 distribution.
Moreover, $m$ scales linearly with $\delta$, therefore the results
we obtain are expected to be quite representative of other parameters sets
as well.

\bigskip
Let us first consider the case of $n=100{,}000$. This indicates that
we are interested in a large-sample behaviour of the estimators.

Table~\ref{tab:biasmse_m} gives the approximate bias and the root mean squared
error of the various estimators of the $m$ parameter.
Recall that for a given estimator $\hat{m}(X_1,\dots,X_n)$
(which is a function of a sequence of random variables)
of the true parameter $m$ (which is a fixed value),
its bias is defined as $\mathbb{E}[\hat{m}(X_1,\dots,X_n)-m]$.
Moreover, the root mean squared error (RMSE) is given by
$
\sqrt{\mathbb{E}[(\hat{m}(X_1,\dots,X_n)-m)^2]}.
$

\begin{table}[hb!]
\caption{\label{tab:biasmse_m}
Bias (and root mean squared error in parentheses)
of different estimators of the $m$ parameter (true $n=100{,}000$, $\delta=1$, $m=25$, and different $\rho$;
$10{,}000$ Monte Carlo samples); the values that the $t$-test deems
not significantly different from $0$ (at the $0.01$ significance level) are greyed out (in theory, the mean is an unbiased estimator of $m$)}

\newcommand{\zero}{\color{gray}}

\begin{tabular}{crrrrrr}
\toprule
        & \multicolumn{2}{c}{$\rho=0.50$}& \multicolumn{2}{c}{$\rho=0.75$}& \multicolumn{2}{c}{$\rho=0.90$}\\
\midrule
\multicolumn{7}{l}{\it continuous data}\\
\midrule
mean    &\zero$-0.00$ & $(0.27)$ & \zero$-0.12$ & $(5.86)$ &      $-1.67$ & $(47.10)$  \\
SSE     &\zero$ 0.01$ & $(0.25)$ &      $ 0.02$ & $(0.58)$ &      $ 0.10$ & $(1.36)$   \\
SUP     &\zero$ 0.01$ & $(0.29)$ &      $ 0.03$ & $(0.65)$ &      $ 0.11$ & $(1.51)$   \\
MLE     &\zero$ 0.00$ & $(0.18)$ &      $ 0.02$ & $(0.41)$ &      $ 0.06$ & $(0.99)$   \\
\midrule
\multicolumn{7}{l}{\it discretised data} \\
\midrule
SSE     &\zero$ 0.00$ & $(0.18)$ &      $ 0.02$ & $(0.43)$ &      $ 0.07$ & $(1.07)$   \\
SUP     &     $ 0.01$ & $(0.26)$ &      $ 0.02$ & $(0.56)$ &      $ 0.09$ & $(1.26)$   \\
MLE     &\zero$ 0.00$ & $(0.18)$ & \zero$ 0.01$ & $(0.41)$ &      $ 0.03$ & $(0.99)$   \\
\bottomrule
\end{tabular}
\end{table}

\begin{table}[hb!]
\caption{\label{tab:biasmse_r}
Bias (and root mean squared error in parentheses)
of different estimators of the $\rho$ parameter
(the same simulation set-up as in Table~\ref{tab:biasmse_m})}

\newcommand{\zero}{\color{gray}}

\begin{tabular}{crrrrrr}
\toprule
        & \multicolumn{2}{c}{$\rho=0.50$}& \multicolumn{2}{c}{$\rho=0.75$}& \multicolumn{2}{c}{$\rho=0.90$}\\
\midrule
\multicolumn{7}{l}{\it continuous data}\\
\midrule
SSE     & \zero$ 0.0001$ & $(0.0072)$ & \zero$ 0.0001$ & $(0.0074)$ & \zero$ 0.0001$ & $(0.0065)$   \\
SUP     & \zero$ 0.0001$ & $(0.0081)$ & \zero$ 0.0001$ & $(0.0081)$ & \zero$ 0.0001$ & $(0.0071)$   \\
MLE     & \zero$ 0.0001$ & $(0.0048)$ & \zero$ 0.0001$ & $(0.0052)$ & \zero$ 0.0001$ & $(0.0047)$   \\
\midrule
\multicolumn{7}{l}{\it discretised data} \\
\midrule
SSE     & \zero$ 0.0001$ & $(0.0050)$ & \zero$ 0.0001$ & $(0.0056)$ & \zero$ 0.0001$ & $(0.0052)$   \\
SUP     & \zero$ 0.0001$ & $(0.0072)$ & \zero$ 0.0001$ & $(0.0071)$ & \zero$ 0.0001$ & $(0.0060)$   \\
MLE     & \zero$ 0.0001$ & $(0.0048)$ & \zero$ 0.0000$ & $(0.0052)$ & \zero$-0.0001$ & $(0.0048)$   \\
\bottomrule
\end{tabular}
\end{table}

We estimate the bias and RMSE based on $M=10{,}000$
(for such a large sample, the computations took a few hours to compute
on a modern PC) independent Monte Carlo samples of size $n$
like $x_1^{(i)},\dots,x_n^{(i)}$, $i=1,\dots,M$, using, respectively,
the sample mean and the standard deviation of a vector
$(e_1,\dots,e_M)$ with $e_i=\hat{m}(x_1^{(i)},\dots,x_n^{(i)})-m$.
In other words, we generate $M$ pseudorandom vectors of size $n$
from the Pareto-type 2 distribution with parameters $m, \rho, \delta$,
estimate the model parameters for each data vector,
compare the estimates to the true values, and then aggregate the results.

The one sample Student $t$-test with a significance level of $0.01$
indicates that for smaller $\rho$s, the $e_i$ values are, on average, not
significantly different from $0$.
This indicates that the estimators might be asymptotically unbiased
(note again the large $n$ we use).

It might be tempting to use the methods of moments
estimator for the $m$ parameter, i.e., the arithmetic mean.
Even though it is an unbiased estimator,
it unfortunately tends to have very high variance
and hence it is a practically useless measure.
We note that $\alpha > 1$ guarantees the existence of the expected value
(which always holds in our case), but the variance is only defined
if $\alpha > 2$, which for $m=25$ and $\delta=1$ holds
whenever $\rho < 25/48 \simeq 0.521$
(but it is not the case in our empirical study in Section~\ref{sec:data}).

Both in the continuous and in the discretised case,
the MLE estimators work very well.
The discretised case seems even more well-behaving.
The MLE estimator should definitely be chosen if we suspect
that data might really come from the distribution studied herein.
Nevertheless, any deviations from the model (such as data contamination)
might affect its performance.
In such a case, the SSE estimator could also be noteworthy.
SUP, on the other hand, is in theory a consistent (i.e., with
asymptotic guarantees; compare the Glivenko--Cantelli theorem)
estimator, but in our case it has a higher root mean squared error.

\begin{table}[t!]
\caption{\label{tab:biasmse_m2}
Bias (and root mean squared error in parentheses)
of different estimators of the $m$ parameter
(true $n=1{,}000$, $\delta=1$, $m=25$, and different $\rho$;
$10{,}000$ Monte Carlo samples)}

\newcommand{\zero}{\color{gray}}

\begin{tabular}{crrrrrr}
\toprule
        & \multicolumn{2}{c}{$\rho=0.50$}& \multicolumn{2}{c}{$\rho=0.75$}& \multicolumn{2}{c}{$\rho=0.90$}\\
\midrule
\multicolumn{7}{l}{\it continuous data}\\
\midrule
mean    & \zero$ 0.00$ & $(2.86)$ & \zero$ 0.05$ & $(49.53)$ & \zero$-0.62$ & $(269.32)$                  \\
SSE     &      $ 0.28$ & $(2.67)$ &      $ 1.97$ & $(9.89)$ &      $ 16.51$ & $(50.57)$                   \\
SUP     &      $ 0.39$ & $(3.06)$ &      $ 2.51$ & $(10.90)$ &      $ 11.16$ & $(31.17)$                  \\
MLE     &      $ 0.11$ & $(1.80)$ &      $ 0.80$ & $(4.91)$ & \zero$ 0.57$ & $(1292.72)$                  \\
\midrule
\multicolumn{7}{l}{\it discretised data} \\
\midrule
SSE      &      $ 0.13$ & $(2.01)$ &      $ 0.90$ & $(5.61)$ &      $ 10.29$ & $(39.31)$           \\
SUP      &      $ 0.31$ & $(2.72)$ &      $ 1.80$ & $(8.51)$ &      $ 10.87$ & $(31.69)$           \\
MLE      &      $ 0.11$ & $(1.80)$ &      $ 0.80$ & $(4.91)$ &      $ 7.39$ & $(27.40)$            \\
\bottomrule
\end{tabular}
\end{table}

\begin{table}[t!]
\caption{\label{tab:biasmse_r2}
Bias (and root mean squared error in parentheses)
of different estimators of the $\rho$ parameter
(the same simulation set-up as in Table~\ref{tab:biasmse_m2})}

\newcommand{\zero}{\color{gray}}

\begin{tabular}{crrrrrr}
\toprule
        & \multicolumn{2}{c}{$\rho=0.50$}& \multicolumn{2}{c}{$\rho=0.75$}& \multicolumn{2}{c}{$\rho=0.90$}\\
\midrule
\multicolumn{7}{l}{\it continuous data}\\
\midrule
SSE     & \zero$-0.0012$ & $(0.0711)$ &      $-0.0020$ & $(0.0721)$ &      $-0.0055$ & $(0.0600)$         \\
SUP     & \zero$-0.0011$ & $(0.0798)$ &      $-0.0025$ & $(0.0795)$ &      $-0.0087$ & $(0.0633)$         \\
MLE     &      $-0.0019$ & $(0.0475)$ &      $-0.0019$ & $(0.0517)$ &      $-0.0025$ & $(0.0469)$         \\
\midrule
\multicolumn{7}{l}{\it discretised data} \\
\midrule
SSE     &      $-0.0030$ & $(0.0546)$ &      $-0.0037$ & $(0.0574)$ &      $-0.0052$ & $(0.0503)$         \\
SUP     &      $-0.0022$ & $(0.0721)$ &      $-0.0027$ & $(0.0705)$ &      $-0.0053$ & $(0.0564)$         \\
MLE     &      $-0.0019$ & $(0.0475)$ &      $-0.0020$ & $(0.0518)$ &      $-0.0033$ & $(0.0461)$         \\
\bottomrule
\end{tabular}
\end{table}

\bigskip
Table~\ref{tab:biasmse_r} provides the approximations to
the bias and root mean squared error of the different estimators of $\rho$.
In all cases, the bias is close to $0$ and the root mean squared
error is quite low.

\bigskip
Unfortunately, for smaller $n$s, the quality of the estimators
deteriorates, particularly in the case of the $m$ parameter,
see Table~\ref{tab:biasmse_m2} and Table~\ref{tab:biasmse_r2}.
In the case of larger $\rho$,
the SUP estimator of $m$ actually becomes quite trustworthy.

\bigskip
We should notice that there are of course many other estimators
of $\alpha$ and $\lambda$ (which we could use for estimating $m$ and $\rho$)
for the continuous case known in the literature, e.g.,
\cite{ZhangStevens2009:estgpd,Luceno2006:estgpd,ZeaKots2001:estgpd,%
Zhang2010:estgpd}.
Many of them are better than MLE for samples of smaller sizes.
However, to the best of our knowledge, there are no discrete counterparts
thereof. This is why we restrict ourselves only to the three above methods.
By the time better methods are developed (which is left for further
research), we suggest that for small sample sizes, the estimated values
should be used with caution.

\section{Analysis of empirical data}\label{sec:data}

Let us fit our model to an example data set.
We shall study the DBLP v12 bibliography database \cite{ArnetMiner}
which covers over $4{,}894{,}081$ papers and
$45{,}564{,}149$ citation relationships.
DBLP includes the most important outlets related to computer science.
As the data consist of natural numbers, we will only fit the discretised
versions of the MLE, SSE, and SUP estimators.

%
%
%
%

\begin{figure}[p!]
\includegraphics[width=\textwidth]{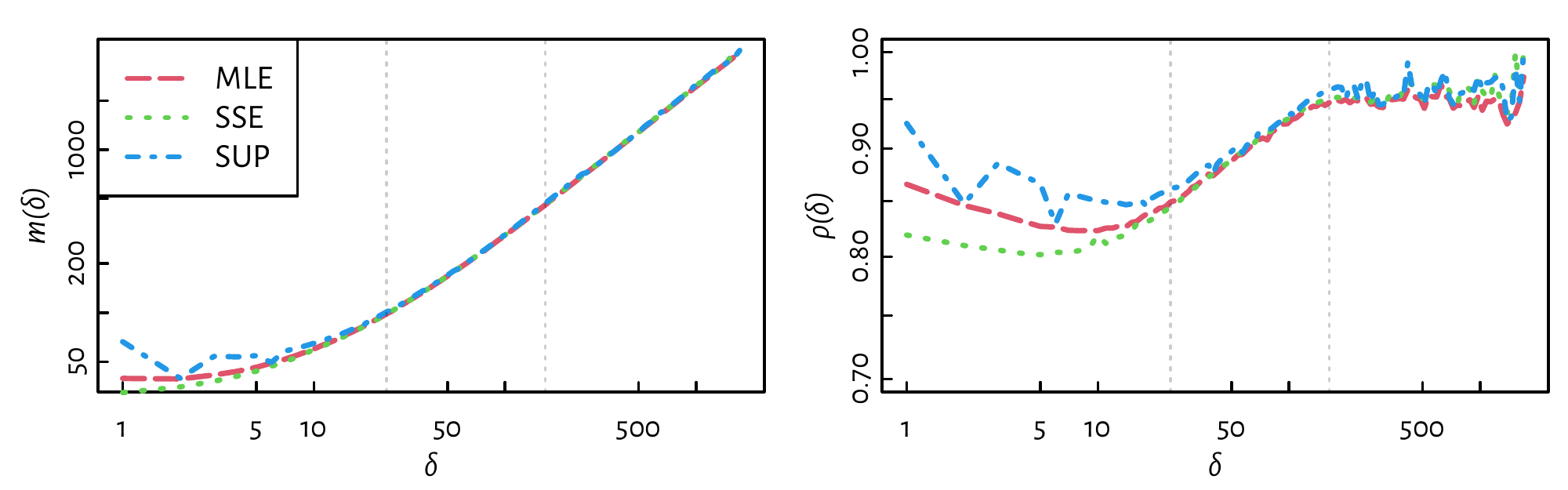}
\caption{\label{Fig:mr_plot}
Estimated parameters $m$ (left), $\rho$ (right)
as functions of the cut-off threshold $\delta$
for DBLP data and different estimators.
Minimal threshold for which the hypothesis of Paretianity is not rejected
is equal to $\delta=20$ (for the SUP estimator at significance level $\alpha=1\%$).
For $\delta\ge 162$, the estimated $\rho$s become degenerated,
but this is where fixing $\rho=1$ (pure power law) leads to a good fit.
}
\end{figure}

\bigskip
Figure~\ref{Fig:mr_plot} depicts how the estimated $m$ and $\rho$
depends on the cutoff threshold $\delta$.

First, let us note that for $\delta<20$,
the discrete Kolmogorov--Smirnov test
rejects the hypothesis of Paretianity for all the estimated parameter pairs
(at significance level of $1\%$). The SUP estimator uses the same
criterion as this goodness-of-fit test, hence this is the method
that recognises some Pareto-type 2 distribution the fastest.
For MLE, we accept the Paretianity hypothesis for all $\delta\ge 29$,
and in the case of SSE, this happens for all $\delta\ge 39$.

Furthermore, for all $\delta\ge 162$, the SUP-based fitting
of the Pareto-type 1 model (pure power law, i.e., with $\rho=1$)
becomes legitimised by the Kolmogorov--Smirnov test.
This is also where the quality of our $\rho$ estimators deteriorates,
which is of course an expected behaviour, because here
the simpler model becomes applicable.

Furthermore, as $\delta$ increases, the size of the sample
drastically decreases. For $\delta=20$, we have $n=803{,}721$.
For $\delta=100$, it holds $n=135{,}982$.
And for $\delta=1{,}000$, we already get $n=4{,}437$.
This also affects the quality of the obtained estimates.

To conclude, our model enables a good description of the
data over a large part of the domain, namely,
for $\delta\in[\sim20, \sim162]$.
We note that all the three estimators yield quite similar parameter values,
which might indicate that they are well behaving in this interval.
The behaviour of $m$ and $\rho$ as a function of the cutoff threshold is
consistent with what we predict from Eq.~\eqref{eq:recreatedthreshold}.


%
%
%
%
%

\begin{figure}[p!]
\includegraphics[width=\textwidth]{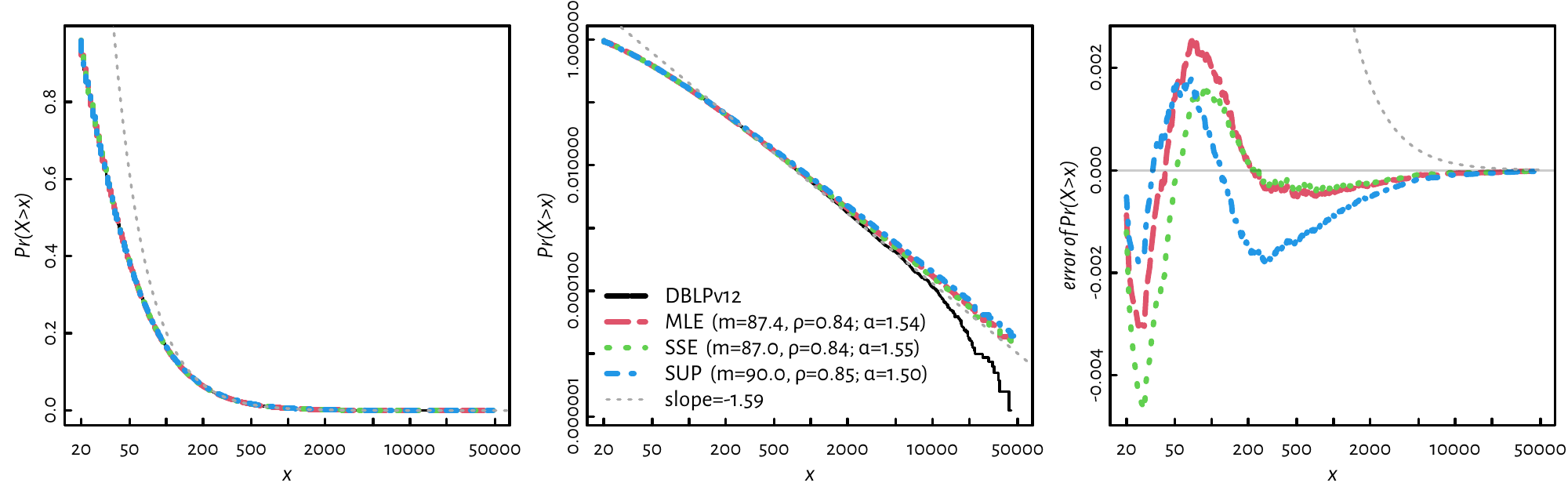}
\caption{\label{Fig:eccdf20}
Empirical (DBLP data, truncated at $\delta=20$)
and fitted complementary cumulative distribution functions;
left: log scale on $Ox$, middle: log scale on both axes,
right: error between the empirical and the theoretical CCDFs;
the power law model (straight line on the log-log scale)
gives a particularly bad fit for smaller observations (which are
the most prevalent);
MLE, SSE, and SUP seem to give poorer fits at the distribution
tail in the middle plot but note that
the probabilities therein are on the log scale,
which exaggerates how the error magnitudes are perceived
}
\end{figure}

\begin{figure}[p!]
\includegraphics[width=\textwidth]{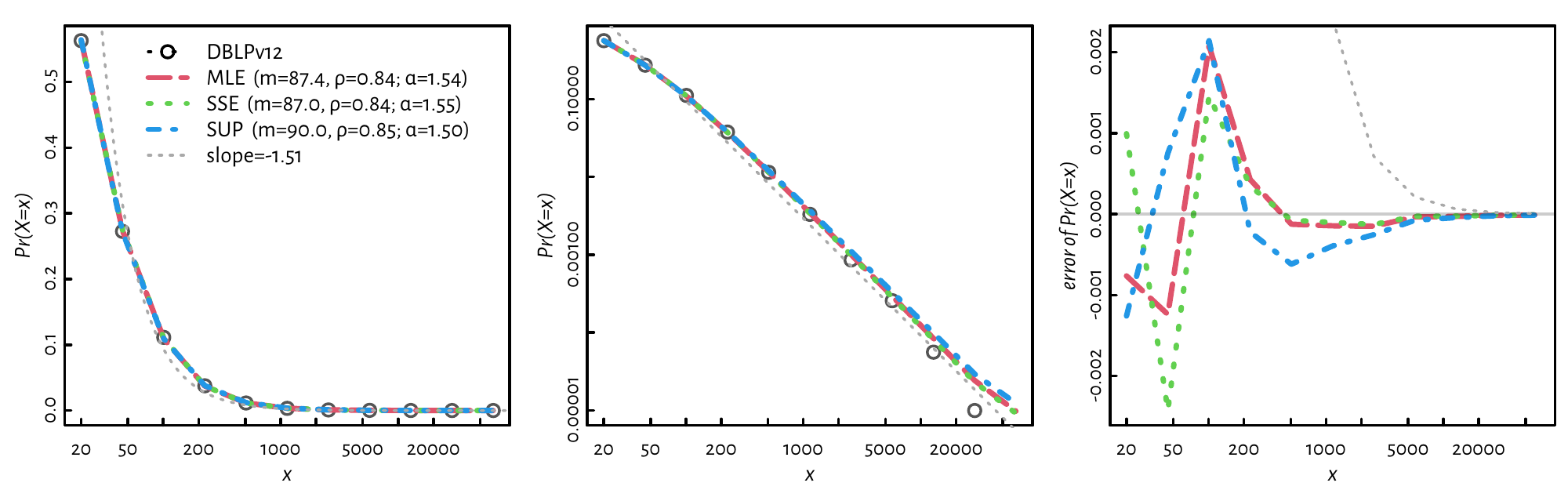}
\caption{\label{Fig:histogram20}
Histogram (logarithmically-binned DBLP data, truncated at $\delta=20$)
and the fitted probability mass functions;
left: log scale on $Ox$, middle: log scale on both axes,
right: error between the histogram and the theoretical PMFs;
note again how the power law model overestimates the density at smaller $x$s
}
\end{figure}

\bigskip
Let us take a closer look at the fitted distributions in the $\delta=20$ case
(average number of citations equal to 82.97).
Figure~\ref{Fig:eccdf20} presents the CCDFs
and Figure~\ref{Fig:histogram20} depicts the logarithmically binned
histogram as well as the fitted probability mass functions.
In terms of the CCDF reproduction accuracy,
the quality of the MLE-fitted model ($m=87.3534102, \rho=0.8406532$)
is somewhere in-between the
SUP ($m=90.0302823, \rho=0.8545986$) and the
SSE ($m=87.0142624, \rho=0.8354341$)
ones.
The estimated parameter values are similar, though.

Note that an inexperienced eye might be tempted to conclude that
in the middle subfigure of Figure \ref{Fig:eccdf20},
the models are not fitted well.
However,
we should emphasise that this graph utilises the double log-scale.
Hence, the impressions of error magnitudes in the tails are exaggerated.
Notably, even for simulated data (that come from the true distribution),
such order statistics (close to the maximum) are subject to high variability.
Therefore, they will from time to time deviate considerably
from the theoretical curve.
It seems, though, that our (asymptotic) model expects the most highly cited
papers in the (finite) sample to be cited more frequently than they really are.

%
%
%
%

\section{Conclusion}\label{sec:final}

Our rank distribution approach marries the Price model
and Pareto-type 2 distributions.
We have shown how a combination of the rich-get-richer rule and sheer chance
yields, in the long run, a well-known statistical distribution
with a new, interpretable parametrisation.
Reversely, the expected values of order statistics of any finite
i.i.d.~Paretian sample are consistent with the baseline Price ranks.

The considered model fits the DBLP citation data
quite well. In fact, the Pareto distribution and its generalisations
have been used in bibliometric modelling a few times already;
see, e.g., \cite{GLANZEL1,GLANZEL2,%
Gagolewski2012:smps,Gagolewski2015:hconfint}.

Interestingly, in our case, for $\delta=20$,
the estimated $\rho$ parameter ($\hat{\rho}_\text{MLE}=0.84$)
corresponds to a very high fraction of
preferentially-attached citations -- other studies of different
real-world databases (e.g., \cite{arnold2021likelihood,peterson2010nonuniversal})
suggested that success might be more accidental.

We noted that for $\rho=1$, our model reduces to the ordinary power law
(the Pareto-type 1 distribution) and that a $\delta$-truncated
Pareto-type 2 distribution yields $\rho\to 1$ as $\delta\to\infty$.
Therefore, we expect that for all the power law-like datasets (e.g., considered
in \cite{Newman2005,ClausetShaliziNewman2009}) we enjoy a fit of the same
or better quality, because we can always truncate at $\delta$
which yields a good power law-fit and assume $\rho=1$.
Still, we leave the empirical study of data from different domains
(e.g., data from linguistics, economics, cellular biology,
communication networks, ecology, transport, modelling of extreme weather
events) for further research.

Future work will also involve the construction of usable estimators of $m$
and $\rho$ for small, discretised samples.
Another interesting research direction is the analysis of the out-degrees
of the network nodes in the Price model,
since in this paper we have only focused on the in-degrees.


\section*{Conflict of interest}

All authors certify that they have no affiliations with or involvement in any
organisation or entity with any financial interest or non-financial interest
in the subject matter or materials discussed in this manuscript.


\section*{Acknowledgement}

This research was supported by the Australian Research Council Discovery
Project ARC DP210100227 (MG) and by the POB Research Centre
Cybersecurity and Data Science of Warsaw University of Technology within
the Excellence Initiative Program -- Research University (ID-UB) (GS and PN).

The authors would also like to thank for valuable remarks the three anonymous
reviewers as well as Anna Cena, Maciej J.~Mrowiński,
and Barbara \.{Z}oga\l{}a-Siudem.


\printcredits

\bibliographystyle{cas-model2-names}

\begin{thebibliography}{49}
\expandafter\ifx\csname natexlab\endcsname\relax\def\natexlab#1{#1}\fi
\providecommand{\url}[1]{\texttt{#1}}
\providecommand{\href}[2]{#2}
\providecommand{\path}[1]{#1}
\providecommand{\DOIprefix}{doi:}
\providecommand{\ArXivprefix}{arXiv:}
\providecommand{\URLprefix}{URL: }
\providecommand{\Pubmedprefix}{pmid:}
\providecommand{\doi}[1]{\href{http://dx.doi.org/#1}{\path{#1}}}
\providecommand{\Pubmed}[1]{\href{pmid:#1}{\path{#1}}}
\providecommand{\bibinfo}[2]{#2}
\ifx\xfnm\relax \def\xfnm[#1]{\unskip,\space#1}\fi
\bibitem[{Arnold(2015)}]{arnold2015:pareto}
\bibinfo{author}{Arnold, B.C.}, \bibinfo{year}{2015}.
\newblock \bibinfo{title}{Pareto Distributions}.
\newblock \bibinfo{publisher}{Chapman and Hall/CRC}, \bibinfo{address}{New
  York, NY, USA}.
\newblock \DOIprefix\doi{10.1201/b18141}.
\bibitem[{Arnold et~al.(2021)Arnold, Mondrag{\'o}n and
  Clegg}]{arnold2021likelihood}
\bibinfo{author}{Arnold, N.A.}, \bibinfo{author}{Mondrag{\'o}n, R.J.},
  \bibinfo{author}{Clegg, R.G.}, \bibinfo{year}{2021}.
\newblock \bibinfo{title}{Likelihood-based approach to discriminate mixtures of
  network models that vary in time}.
\newblock \bibinfo{journal}{Scientific Reports} \bibinfo{volume}{11},
  \bibinfo{pages}{1--13}.
\bibitem[{Bedogne and Rodgers(2006)}]{bedogne2006complex}
\bibinfo{author}{Bedogne, C.}, \bibinfo{author}{Rodgers, G.J.},
  \bibinfo{year}{2006}.
\newblock \bibinfo{title}{Complex growing networks with intrinsic vertex
  fitness}.
\newblock \bibinfo{journal}{Physical Review E} \bibinfo{volume}{74},
  \bibinfo{pages}{046115}.
\bibitem[{Beirlant et~al.(2004)Beirlant, Goegebeur, Teugels and Segers}]{evt}
\bibinfo{author}{Beirlant, J.}, \bibinfo{author}{Goegebeur, Y.},
  \bibinfo{author}{Teugels, J.}, \bibinfo{author}{Segers, J.},
  \bibinfo{year}{2004}.
\newblock \bibinfo{title}{Statistics of Extremes: {T}heory and Applications}.
\newblock \bibinfo{publisher}{Wiley}.
\newblock \DOIprefix\doi{10.1002/0470012382}.
\bibitem[{Bermudez and Kotz(2010)}]{ZeaKots2001:estgpd}
\bibinfo{author}{Bermudez, P.Z.}, \bibinfo{author}{Kotz, S.},
  \bibinfo{year}{2010}.
\newblock \bibinfo{title}{Parameter estimation of the {G}eneralized {P}areto
  {D}istribution. {P}art {II}}.
\newblock \bibinfo{journal}{Journal of Statistical Planning and Inference}
  \bibinfo{volume}{140}, \bibinfo{pages}{1374--1388}.
\bibitem[{Billingsley(1995)}]{Billingsley}
\bibinfo{author}{Billingsley, P.}, \bibinfo{year}{1995}.
\newblock \bibinfo{title}{Probability and Measure}.
\newblock \bibinfo{publisher}{Wiley}.
\bibitem[{Bourguignon et~al.(2016)Bourguignon, Saulo and
  Fernandez}]{bourguignon2016new}
\bibinfo{author}{Bourguignon, M.}, \bibinfo{author}{Saulo, H.},
  \bibinfo{author}{Fernandez, R.N.}, \bibinfo{year}{2016}.
\newblock \bibinfo{title}{A new pareto-type distribution with applications in
  reliability and income data}.
\newblock \bibinfo{journal}{Physica A: Statistical Mechanics and its
  Applications} \bibinfo{volume}{457}, \bibinfo{pages}{166--175}.
\bibitem[{Clauset et~al.(2009)Clauset, Shalizi and
  Newman}]{ClausetShaliziNewman2009}
\bibinfo{author}{Clauset, A.}, \bibinfo{author}{Shalizi, C.R.},
  \bibinfo{author}{Newman, M.E.}, \bibinfo{year}{2009}.
\newblock \bibinfo{title}{Power-law distributions in empirical data}.
\newblock \bibinfo{journal}{SIAM Review} \bibinfo{volume}{51},
  \bibinfo{pages}{661--703}.
\bibitem[{David and Nagaraja(2003)}]{DavidNagaraja2003:orderstatistics}
\bibinfo{author}{David, H.A.}, \bibinfo{author}{Nagaraja, H.N.},
  \bibinfo{year}{2003}.
\newblock \bibinfo{title}{Order statistics}.
\newblock \bibinfo{publisher}{Wiley}.
\bibitem[{Dorogovtsev and Mendes(2015)}]{dorogovtsev2015ranking}
\bibinfo{author}{Dorogovtsev, S.N.}, \bibinfo{author}{Mendes, J.F.},
  \bibinfo{year}{2015}.
\newblock \bibinfo{title}{Ranking scientists}.
\newblock \bibinfo{journal}{Nature Physics} \bibinfo{volume}{11},
  \bibinfo{pages}{882--883}.
\bibitem[{Dorogovtsev et~al.(2000)Dorogovtsev, Mendes and
  Samukhin}]{Dorogovtsev2000}
\bibinfo{author}{Dorogovtsev, S.N.}, \bibinfo{author}{Mendes, J.F.F.},
  \bibinfo{author}{Samukhin, A.N.}, \bibinfo{year}{2000}.
\newblock \bibinfo{title}{Structure of growing networks with preferential
  linking}.
\newblock \bibinfo{journal}{Phys. Rev. Lett.} \bibinfo{volume}{85},
  \bibinfo{pages}{4633--4636}.
\newblock \DOIprefix\doi{10.1103/PhysRevLett.85.4633}.
\bibitem[{Evans et~al.(2020)Evans, Calmon and
  Vasiliauskaite}]{evans2020longest}
\bibinfo{author}{Evans, T.S.}, \bibinfo{author}{Calmon, L.},
  \bibinfo{author}{Vasiliauskaite, V.}, \bibinfo{year}{2020}.
\newblock \bibinfo{title}{The longest path in the {P}rice model}.
\newblock \bibinfo{journal}{Scientific Reports} \bibinfo{volume}{10},
  \bibinfo{pages}{1--9}.
\bibitem[{Figueira et~al.(2011)Figueira, Moura~Jr and
  Ribeiro}]{figueira2011gompertz}
\bibinfo{author}{Figueira, F.C.}, \bibinfo{author}{Moura~Jr, N.J.},
  \bibinfo{author}{Ribeiro, M.B.}, \bibinfo{year}{2011}.
\newblock \bibinfo{title}{The gompertz--pareto income distribution}.
\newblock \bibinfo{journal}{Physica A: Statistical Mechanics and its
  Applications} \bibinfo{volume}{390}, \bibinfo{pages}{689--698}.
\bibitem[{Gagolewski(2013a)}]{Gagolewski2013:fair}
\bibinfo{author}{Gagolewski, M.}, \bibinfo{year}{2013}a.
\newblock \bibinfo{title}{Scientific impact assessment cannot be fair}.
\newblock \bibinfo{journal}{Journal of Informetrics} \bibinfo{volume}{7},
  \bibinfo{pages}{792--802}.
\bibitem[{Gagolewski(2013b)}]{Gagolewski2012:smps}
\bibinfo{author}{Gagolewski, M.}, \bibinfo{year}{2013}b.
\newblock \bibinfo{title}{Statistical hypothesis test for the difference
  between {H}irsch indices of two {P}areto-distributed random samples}, in:
  \bibinfo{editor}{Kruse, R.}, et~al. (Eds.), \bibinfo{booktitle}{Synergies of
  Soft Computing and Statistics for Intelligent Data Analysis}.
  \bibinfo{publisher}{Springer}. volume \bibinfo{volume}{190} of
  \textit{\bibinfo{series}{Advances in Intelligent Systems and Computing}}, pp.
  \bibinfo{pages}{359--367}.
\newblock \DOIprefix\doi{10.1007/978-3-642-33042-1_39}.
\bibitem[{Gagolewski(2015)}]{Gagolewski2015:hconfint}
\bibinfo{author}{Gagolewski, M.}, \bibinfo{year}{2015}.
\newblock \bibinfo{title}{Sugeno integral-based confidence intervals for the
  theoretical h-index}, in: \bibinfo{editor}{Grzegorzewski, P.}, et~al. (Eds.),
  \bibinfo{booktitle}{Strengthening Links Between Data Analysis and Soft
  Computing}. \bibinfo{publisher}{Springer}. volume \bibinfo{volume}{315} of
  \textit{\bibinfo{series}{Advances in Intelligent Systems and Computing}}, pp.
  \bibinfo{pages}{233--240}.
\newblock \DOIprefix\doi{10.1007/978-3-319-10765-3_28}.
\bibitem[{Gautschi(1959)}]{Gautschi}
\bibinfo{author}{Gautschi, W.}, \bibinfo{year}{1959}.
\newblock \bibinfo{title}{Some elementary inequalities relating to the gamma
  and incomplete gamma function}.
\newblock \bibinfo{journal}{Journal of Mathematics and Physics}
  \bibinfo{volume}{38}, \bibinfo{pages}{77--81}.
\newblock \DOIprefix\doi{10.1002/sapm195938177}.
\bibitem[{Glänzel(2006)}]{GLANZEL2}
\bibinfo{author}{Glänzel, W.}, \bibinfo{year}{2006}.
\newblock \bibinfo{title}{On the h-index -- {A} mathematical approach to a new
  measure of publication activity and citation impact}.
\newblock \bibinfo{journal}{Scientometrics} \bibinfo{volume}{67},
  \bibinfo{pages}{315--321}.
\bibitem[{Golosovsky(2018)}]{Golosovsky2018}
\bibinfo{author}{Golosovsky, M.}, \bibinfo{year}{2018}.
\newblock \bibinfo{title}{Mechanisms of complex network growth: Synthesis of
  the preferential attachment and fitness models}.
\newblock \bibinfo{journal}{Phys. Rev. E} \bibinfo{volume}{97},
  \bibinfo{pages}{062310}.
\newblock \DOIprefix\doi{10.1103/PhysRevE.97.062310}.
\bibitem[{Golosovsky and Solomon(2017)}]{Golosovsky2017}
\bibinfo{author}{Golosovsky, M.}, \bibinfo{author}{Solomon, S.},
  \bibinfo{year}{2017}.
\newblock \bibinfo{title}{Growing complex network of citations of scientific
  papers: Modeling and measurements}.
\newblock \bibinfo{journal}{Phys. Rev. E} \bibinfo{volume}{95},
  \bibinfo{pages}{012324}.
\newblock \DOIprefix\doi{10.1103/PhysRevE.95.012324}.
\bibitem[{Heesen(2017)}]{heesen2017academic}
\bibinfo{author}{Heesen, R.}, \bibinfo{year}{2017}.
\newblock \bibinfo{title}{Academic superstars: {C}ompetent or lucky?}
\newblock \bibinfo{journal}{Synthese} \bibinfo{volume}{194},
  \bibinfo{pages}{4499--4518}.
\bibitem[{Ionescu and Chopard(2013)}]{IonescuChopard}
\bibinfo{author}{Ionescu, G.}, \bibinfo{author}{Chopard, B.},
  \bibinfo{year}{2013}.
\newblock \bibinfo{title}{An agent-based model for the bibliometric h-index}.
\newblock \bibinfo{journal}{Eur. Phys. J. B} \bibinfo{volume}{86},
  \bibinfo{pages}{426}.
\bibitem[{Janosov et~al.(2020)Janosov, Battiston and
  Sinatra}]{janosov2020success}
\bibinfo{author}{Janosov, M.}, \bibinfo{author}{Battiston, F.},
  \bibinfo{author}{Sinatra, R.}, \bibinfo{year}{2020}.
\newblock \bibinfo{title}{Success and luck in creative careers}.
\newblock \bibinfo{journal}{EPJ Data Science} \bibinfo{volume}{9},
  \bibinfo{pages}{9}.
\bibitem[{Kharel et~al.(2021)Kharel, Mezei, Chung, Erd{\H{o}}s and
  Toroczkai}]{kharel2021degree}
\bibinfo{author}{Kharel, S.R.}, \bibinfo{author}{Mezei, T.R.},
  \bibinfo{author}{Chung, S.}, \bibinfo{author}{Erd{\H{o}}s, P.L.},
  \bibinfo{author}{Toroczkai, Z.}, \bibinfo{year}{2021}.
\newblock \bibinfo{title}{Degree-preserving network growth}.
\newblock \bibinfo{journal}{Nature Physics} , \bibinfo{pages}{1--7}.
\bibitem[{Liu et~al.(2002)Liu, Lai, Ye and Dasgupta}]{liu2002connectivity}
\bibinfo{author}{Liu, Z.}, \bibinfo{author}{Lai, Y.C.}, \bibinfo{author}{Ye,
  N.}, \bibinfo{author}{Dasgupta, P.}, \bibinfo{year}{2002}.
\newblock \bibinfo{title}{Connectivity distribution and attack tolerance of
  general networks with both preferential and random attachments}.
\newblock \bibinfo{journal}{Physics Letters A} \bibinfo{volume}{303},
  \bibinfo{pages}{337--344}.
\bibitem[{Luceno(2006)}]{Luceno2006:estgpd}
\bibinfo{author}{Luceno, A.}, \bibinfo{year}{2006}.
\newblock \bibinfo{title}{Fitting the {G}eneralized {P}areto {D}istribution to
  data using maximum goodness-of-fit estimators}.
\newblock \bibinfo{journal}{Computational Statistics and Data Analysis}
  \bibinfo{volume}{1}, \bibinfo{pages}{904--917}.
\bibitem[{Lyu et~al.(2021)Lyu, Ruan, Xie and Cheng}]{lyu2021classification}
\bibinfo{author}{Lyu, D.}, \bibinfo{author}{Ruan, X.}, \bibinfo{author}{Xie,
  J.}, \bibinfo{author}{Cheng, Y.}, \bibinfo{year}{2021}.
\newblock \bibinfo{title}{The classification of citing motivations: {A}
  meta-synthesis}.
\newblock \bibinfo{journal}{Scientometrics} \bibinfo{volume}{126},
  \bibinfo{pages}{3243--3264}.
\bibitem[{Merton(1968)}]{Merton1968}
\bibinfo{author}{Merton, R.K.}, \bibinfo{year}{1968}.
\newblock \bibinfo{title}{The {M}atthew effect in science}.
\newblock \bibinfo{journal}{Science} \bibinfo{volume}{159},
  \bibinfo{pages}{56--63}.
\newblock \DOIprefix\doi{10.1126/science.159.3810.56}.
\bibitem[{N\'{e}da et~al.(2017)N\'{e}da, Varga and Bir\'{o}}]{Neda2017}
\bibinfo{author}{N\'{e}da, Z.}, \bibinfo{author}{Varga, L.},
  \bibinfo{author}{Bir\'{o}, T.S.}, \bibinfo{year}{2017}.
\newblock \bibinfo{title}{Science and {F}acebook: {T}he same popularity law!}
\newblock \bibinfo{journal}{PLOS ONE} \bibinfo{volume}{12},
  \bibinfo{pages}{1--11}.
\newblock \DOIprefix\doi{10.1371/journal.pone.0179656}.
\bibitem[{Newman(2018)}]{newman2018networks}
\bibinfo{author}{Newman, M.}, \bibinfo{year}{2018}.
\newblock \bibinfo{title}{Networks}.
\newblock \bibinfo{publisher}{Oxford University Press}.
\bibitem[{Newman(2005)}]{Newman2005}
\bibinfo{author}{Newman, M.E.}, \bibinfo{year}{2005}.
\newblock \bibinfo{title}{Power laws, {P}areto distributions and {Z}ipf's law}.
\newblock \bibinfo{journal}{Contemporary Physics} \bibinfo{volume}{46},
  \bibinfo{pages}{323--351}.
\newblock \DOIprefix\doi{10.1080/00107510500052444}.
\bibitem[{Perc(2014)}]{perc2014matthew}
\bibinfo{author}{Perc, M.}, \bibinfo{year}{2014}.
\newblock \bibinfo{title}{The {M}atthew effect in empirical data}.
\newblock \bibinfo{journal}{Journal of The Royal Society Interface}
  \bibinfo{volume}{11}, \bibinfo{pages}{20140378}.
\bibitem[{Peterson et~al.(2010)Peterson, Press{\'e} and
  Dill}]{peterson2010nonuniversal}
\bibinfo{author}{Peterson, G.J.}, \bibinfo{author}{Press{\'e}, S.},
  \bibinfo{author}{Dill, K.A.}, \bibinfo{year}{2010}.
\newblock \bibinfo{title}{Nonuniversal power law scaling in the probability
  distribution of scientific citations}.
\newblock \bibinfo{journal}{Proceedings of the National Academy of Sciences}
  \bibinfo{volume}{107}, \bibinfo{pages}{16023--16027}.
\bibitem[{Pluchino et~al.(2018)Pluchino, Biondo and
  Rapisarda}]{pluchino2018talent}
\bibinfo{author}{Pluchino, A.}, \bibinfo{author}{Biondo, A.E.},
  \bibinfo{author}{Rapisarda, A.}, \bibinfo{year}{2018}.
\newblock \bibinfo{title}{Talent versus luck: {T}he role of randomness in
  success and failure}.
\newblock \bibinfo{journal}{Advances in Complex systems} \bibinfo{volume}{21},
  \bibinfo{pages}{1850014}.
\bibitem[{Pluchino et~al.(2019)Pluchino, Burgio, Rapisarda, Biondo, Pulvirenti,
  Ferro and Giorgino}]{pluchino2019exploring}
\bibinfo{author}{Pluchino, A.}, \bibinfo{author}{Burgio, G.},
  \bibinfo{author}{Rapisarda, A.}, \bibinfo{author}{Biondo, A.E.},
  \bibinfo{author}{Pulvirenti, A.}, \bibinfo{author}{Ferro, A.},
  \bibinfo{author}{Giorgino, T.}, \bibinfo{year}{2019}.
\newblock \bibinfo{title}{Exploring the role of interdisciplinarity in physics:
  {S}uccess, talent and luck}.
\newblock \bibinfo{journal}{PLOS ONE} \bibinfo{volume}{14},
  \bibinfo{pages}{e0218793}.
\bibitem[{Price(1963)}]{Price1963-book}
\bibinfo{author}{Price, D.}, \bibinfo{year}{1963}.
\newblock \bibinfo{title}{Little Science, Big Science}.
\newblock \bibinfo{publisher}{Columbia Univ. Press}, \bibinfo{address}{New
  York}.
\bibitem[{Price(1976)}]{price1976general}
\bibinfo{author}{Price, D.}, \bibinfo{year}{1976}.
\newblock \bibinfo{title}{A general theory of bibliometric and other cumulative
  advantage processes}.
\newblock \bibinfo{journal}{Journal of the American Society for Information
  Science} \bibinfo{volume}{27}, \bibinfo{pages}{292--306}.
\bibitem[{Schubert and Glänzel(2007)}]{GLANZEL1}
\bibinfo{author}{Schubert, A.}, \bibinfo{author}{Glänzel, W.},
  \bibinfo{year}{2007}.
\newblock \bibinfo{title}{A systematic analysis of {H}irsch-type indices for
  journals}.
\newblock \bibinfo{journal}{Journal of Informetrics} \bibinfo{volume}{1},
  \bibinfo{pages}{179--184}.
\bibitem[{Shao et~al.(2006)Shao, Zou, Tan and Jin}]{shao2006growing}
\bibinfo{author}{Shao, Z.G.}, \bibinfo{author}{Zou, X.W.},
  \bibinfo{author}{Tan, Z.J.}, \bibinfo{author}{Jin, Z.Z.},
  \bibinfo{year}{2006}.
\newblock \bibinfo{title}{Growing networks with mixed attachment mechanisms}.
\newblock \bibinfo{journal}{Journal of Physics A: Mathematical and General}
  \bibinfo{volume}{39}, \bibinfo{pages}{2035}.
\bibitem[{Sinatra et~al.(2016)Sinatra, Wang, Deville, Song and
  Barab{\'a}si}]{sinatra2016quantifying}
\bibinfo{author}{Sinatra, R.}, \bibinfo{author}{Wang, D.},
  \bibinfo{author}{Deville, P.}, \bibinfo{author}{Song, C.},
  \bibinfo{author}{Barab{\'a}si, A.L.}, \bibinfo{year}{2016}.
\newblock \bibinfo{title}{Quantifying the evolution of individual scientific
  impact}.
\newblock \bibinfo{journal}{Science} \bibinfo{volume}{354}.
\bibitem[{Siudem et~al.(2020)Siudem, {Żogała-Siudem}, Cena and
  Gagolewski}]{PNAS2020}
\bibinfo{author}{Siudem, G.}, \bibinfo{author}{{Żogała-Siudem}, B.},
  \bibinfo{author}{Cena, A.}, \bibinfo{author}{Gagolewski, M.},
  \bibinfo{year}{2020}.
\newblock \bibinfo{title}{Three dimensions of scientific impact}.
\newblock \bibinfo{journal}{Proceedings of the National Academy of Sciences}
  \bibinfo{volume}{117}, \bibinfo{pages}{13896--13900}.
\newblock \DOIprefix\doi{10.1073/pnas.2001064117}.
\bibitem[{Steinbock et~al.(2017)Steinbock, Biham and Katzav}]{Steinbock2017}
\bibinfo{author}{Steinbock, C.}, \bibinfo{author}{Biham, O.},
  \bibinfo{author}{Katzav, E.}, \bibinfo{year}{2017}.
\newblock \bibinfo{title}{Distribution of shortest path lengths in a class of
  node duplication network models}.
\newblock \bibinfo{journal}{Phys. Rev. E} \bibinfo{volume}{96},
  \bibinfo{pages}{032301}.
\newblock \DOIprefix\doi{10.1103/PhysRevE.96.032301}.
\bibitem[{Steinbock et~al.(2019a)Steinbock, Biham and
  Katzav}]{steinbock2019analytical}
\bibinfo{author}{Steinbock, C.}, \bibinfo{author}{Biham, O.},
  \bibinfo{author}{Katzav, E.}, \bibinfo{year}{2019}a.
\newblock \bibinfo{title}{Analytical results for the distribution of shortest
  path lengths in directed random networks that grow by node duplication}.
\newblock \bibinfo{journal}{The European Physical Journal B}
  \bibinfo{volume}{92}, \bibinfo{pages}{1--16}.
\bibitem[{Steinbock et~al.(2019b)Steinbock, Biham and
  Katzav}]{steinbock2019analytical2}
\bibinfo{author}{Steinbock, C.}, \bibinfo{author}{Biham, O.},
  \bibinfo{author}{Katzav, E.}, \bibinfo{year}{2019}b.
\newblock \bibinfo{title}{Analytical results for the in-degree and out-degree
  distributions of directed random networks that grow by node duplication}.
\newblock \bibinfo{journal}{Journal of Statistical Mechanics: Theory and
  Experiment} \bibinfo{volume}{2019}, \bibinfo{pages}{083403}.
\bibitem[{Tang et~al.(2008)Tang, Zhang, Yao, Li, Zhang and Su}]{ArnetMiner}
\bibinfo{author}{Tang, J.}, \bibinfo{author}{Zhang, J.}, \bibinfo{author}{Yao,
  L.}, \bibinfo{author}{Li, J.}, \bibinfo{author}{Zhang, L.},
  \bibinfo{author}{Su, Z.}, \bibinfo{year}{2008}.
\newblock \bibinfo{title}{Arnetminer: {E}xtraction and mining of academic
  social networks}, in: \bibinfo{booktitle}{Proceedings of the 14th ACM SIGKDD
  international conference on Knowledge discovery and data mining}, pp.
  \bibinfo{pages}{990--998}.
\bibitem[{Thelwall and Sud(2021)}]{thelwall2021new}
\bibinfo{author}{Thelwall, M.}, \bibinfo{author}{Sud, P.},
  \bibinfo{year}{2021}.
\newblock \bibinfo{title}{Do new research issues attract more citations? {A}
  comparison between 25 {S}copus subject categories}.
\newblock \bibinfo{journal}{Journal of the Association for Information Science
  and Technology} \bibinfo{volume}{72}, \bibinfo{pages}{269--279}.
\bibitem[{Zhang(2010)}]{Zhang2010:estgpd}
\bibinfo{author}{Zhang, J.}, \bibinfo{year}{2010}.
\newblock \bibinfo{title}{Improving on estimation for the {G}eneralized
  {P}areto {D}istribution}.
\newblock \bibinfo{journal}{Technometrics} \bibinfo{volume}{52},
  \bibinfo{pages}{335--339}.
\bibitem[{Zhang and Stephens(2009)}]{ZhangStevens2009:estgpd}
\bibinfo{author}{Zhang, J.}, \bibinfo{author}{Stephens, M.A.},
  \bibinfo{year}{2009}.
\newblock \bibinfo{title}{A new and efficient estimation method for the
  {G}eneralized {P}areto {D}istribution}.
\newblock \bibinfo{journal}{Technometrics} \bibinfo{volume}{51},
  \bibinfo{pages}{316--325}.
\bibitem[{{Żogała-Siudem} et~al.(2016){Żogała-Siudem}, Siudem, Cena and
  Gagolewski}]{Zogala-Siudem2016}
\bibinfo{author}{{Żogała-Siudem}, B.}, \bibinfo{author}{Siudem, G.},
  \bibinfo{author}{Cena, A.}, \bibinfo{author}{Gagolewski, M.},
  \bibinfo{year}{2016}.
\newblock \bibinfo{title}{Agent-based model for the h-index -- {E}xact
  solution}.
\newblock \bibinfo{journal}{European Physical Journal B} , \bibinfo{pages}{21}.

\end{thebibliography}

\end{document}